%% file: bare_jrnl_compsoc.tex
\documentclass[10pt,journal,compsoc]{IEEEtran}

\usepackage{amsmath,amssymb,amsfonts}
\usepackage{graphicx}
\usepackage{textcomp}
\usepackage{xcolor}
\usepackage{tabularx}
\usepackage{booktabs} 
\usepackage{longtable}
\usepackage{tabu}
\usepackage{booktabs} 
\usepackage{algorithm}
\usepackage{algorithmicx}
\usepackage{algpseudocode}
\usepackage{url}
\usepackage{epstopdf}
\usepackage{float}
\newcommand{\ignore}[1]{}

\newtheorem{defn}{{\bf Definition}}
\newtheorem{insight}{{\bf Insight}}
\ifCLASSOPTIONcompsoc
  \usepackage[nocompress]{cite}
\else
  \usepackage{cite}
\fi
\ifCLASSINFOpdf
\else
\fi
\begin{document}

\title{$\mu$VulDeePecker: A Deep Learning-Based System for Multiclass Vulnerability Detection\\}

\author{Deqing~Zou,
	Sujuan~Wang,
	Shouhuai~Xu,
	Zhen~Li,
    and~Hai~Jin,~\IEEEmembership{Fellow,~IEEE}
\thanks{Corresponding author: Deqing Zou.}
\IEEEcompsocitemizethanks{
	\IEEEcompsocthanksitem D. Zou is with National Engineering Research Center for Big Data Technology and System, Services Computing Technology and System Lab, Cluster and Grid Computing Lab, Big Data Security Engineering Research Center, School of Cyber Science and Engineering, Huazhong University of Science and Technology, Wuhan 430074, China, and also with Shenzhen Huazhong University of Science and Technology Research Institute, Shenzhen 518057, China. E-mail: deqingzou@hust.edu.cn.
	\IEEEcompsocthanksitem S. Wang, Z. Li and H. Jin are with National Engineering Research Center for Big Data Technology and System, Services Computing Technology and System Lab, Cluster and Grid Computing Lab, Big Data Security Engineering Research Center, School of Computer
	Science and Technology, Huazhong University of Science and Technology, Wuhan 430074, China. E-mail: \{sophiewsj, lizhen\_hust, hjin\}@hust.edu.cn.
	
\IEEEcompsocthanksitem S. Xu is with the Department of Computer Science, University of Texas at San Antonio, San Antonio TX 78249, USA. E-mall: shxu@cs.utsa.edu.}
}

%
%

\markboth{IEEE Transactions on Dependable and Secure Computing}%
{Shell \MakeLowercase{\textit{et al.}}: Bare Demo of IEEEtran.cls for Computer Society Journals}
%



\IEEEtitleabstractindextext{%
\begin{abstract}
Fine-grained software vulnerability detection is an important and challenging problem. Ideally, a detection system (or detector) not only should be able to detect whether or not a program contains vulnerabilities, but also should be able to pinpoint the type of a vulnerability in question. Existing vulnerability detection methods based on deep learning can detect the presence of vulnerabilities (i.e., addressing the binary classification or detection problem), but cannot pinpoint types of vulnerabilities (i.e., incapable of addressing multiclass classification). In this paper, we propose the first deep learning-based system for multiclass vulnerability detection, dubbed $\mu$VulDeePecker. The key insight underlying $\mu$VulDeePecker is the concept of {\em code attention}, which can capture information that can help pinpoint types of vulnerabilities, even when the samples are small. For this purpose, we create a dataset from scratch and use it to evaluate the effectiveness of $\mu$VulDeePecker. Experimental results show that $\mu$VulDeePecker is effective for multiclass vulnerability detection and that accommodating control-dependence (other than data-dependence) can lead to higher detection capabilities.
\end{abstract}

}

\maketitle

\IEEEdisplaynontitleabstractindextext

%
\IEEEpeerreviewmaketitle

\input{samplebody-conf}

\section*{Acknowledgments}
We thank the anonymous reviewers for their constructive comments that guided us in improving the paper.
The authors at Huazhong University of Science and Technology is supported in part by the National Key Research and Development Plan of China under Grant No.2017YFB0802205, in part by the National Natural Science Foundation of China under Grant No.61672249 and No.61802106, and in part by the Shenzhen Fundamental Research Program under Grant No.JCYJ20170413114215614. 
Shouhuai Xu is supported in part by NSF CREST Grant No. 1736209. The opinions expressed in the paper are those of the authors' and do not reflect the funding agencies' policies in any sense.

\bibliographystyle{IEEEtran}
\bibliography{bare_jrnl_compsoc}

\end{document}

%% file: samplebody-conf.tex
\section{Introduction}
\label{sec:introduction}

\IEEEPARstart{M}{ost} cyber attacks are caused by the exploitation of one or multiple vulnerabilities \cite{Sun2019datadriven,DBLP:journals/comsur/LiuVHZX18}. It is unfortunate that vulnerabilities are inevitable, for many reasons (e.g., software complexity, steady growth in internet complexity\cite{DBLP:journals/chinaf/RajehJZ17}).
Given that vulnerabilities cannot be prevented, an alternate defense approach is to detect and patch vulnerabilities sooner rather than later, leading to the field of {\em vulnerability detection}.
This problem has received a due amount of attention, leading to many approaches. A popular approach is to use {\em manually-defined} patterns to detect vulnerabilities \cite{cx, cppchecker, cvechecker, jang2012redebug, li2012cbcd, yamaguchi2013chucky, sajnani2016sourcerercc}, which
attained a limited success. Another promising approach is to use machine learning for vulnerability detection (see, for example, \cite{yamaguchi2011vulnerability, walden2014predicting}). These solutions reduce the workload on human experts because they only need to roughly define features for learning machine learning-based models that can detect vulnerabilities, rather than defining vulnerability patterns manually. Compared with traditional machine learning techniques, researchers have started using deep learning for detecting vulnerabilities \cite{li2018vuldeepecker, lin2017poster} and software defects \cite{wang2016automatically, dam2018deep}.

The state-of-the-art of deep learning-based vulnerability detection is a system called VulDeePecker \cite{li2018vuldeepecker}, which uses Bidirectional Long-Short Time Memory (BLSTM) neural network to detect software vulnerabilities. However, VulDeePecker is a binary classifier or detector, meaning that it can tell whether a piece of code (i.e., multiple lines of code) is vulnerable or not, but cannot pinpoint the type of the vulnerability in question. The type of a vulnerability is important because this information will tell the vulnerability principles and help quickly pin down the precise location of a vulnerability and reduce the workload of developers and code  auditors, which is especially important when the piece of code has a substantial number of lines (e.g., tens number of lines of code), which is also common to machine learning-based vulnerability detection systems \cite{ghaffarian2017software}. In this paper, we move a step towards ultimately tackling this problem by investigating {\em multiclass} vulnerability detection, which not only can tell whether a piece of code is vulnerable or not, but also can pinpoint the type of a vulnerability.

\subsection{Our Contributions}

We initiate the investigation on {\em multiclass} vulnerability detection and present a system that uses deep learning for this purpose. The system is called \underline{mu}lticlass \underline{Vul}nerability \underline{Dee}p \underline{Pecker}, or $\mu$VulDeePecker for short.

The innovations of the system are in three-fold. First, a conceptual innovation underlying $\mu$VulDeePecker is the introduction of the concept we call {\em code attention}, which can accommodate information useful for learning {\em local features} and helping pinpoint types of vulnerabilities. It refines the concept of {\em code gadget} \cite{li2018vuldeepecker} which is a number of statements that are semantically related to each other. 
Second, another innovation underlying $\mu$VulDeePecker is redefining the concept and extraction method of code gadget by introducing control-dependence. The last innovation underlying $\mu$VulDeePecker is a new neural network architecture. The architecture is different from the models used in previous vulnerability detection methods. It is mainly constructed from building-block BLSTM networks and aims to fuse different kinds of features from {\em code gadget} and {\em code attention} to accommodate different kinds of information. This neural network architecture may be of independent value because it provides an effective fused idea for different code features and can be referenced in other scenarios.
	
In order to evaluate the effectiveness of $\mu$VulDeePecker, we create a dataset that contains 181,641 pieces of code (called code gadgets, which are units for vulnerability detection) from 33,409 programs.
Among them, 138,522 are non-vulnerable (i.e., not known to contain vulnerabilities) and the other 43,119 are vulnerable and contain 40 types of vulnerabilities in total.
This dataset should be useful to other researchers and is available at \url{https://github.com/muVulDeePecker/muVulDeePecker}.
Systematic experiments using this dataset show the following:
(i) $\mu$VulDeePecker is effective for multiclass vulnerability detection. In particular, the use of code attention can indeed help recognize types of vulnerabilities, even if for small samples. It is significant because deep learning often requires large amounts of data in order to be effective.
(ii) The accommodation of control-dependence can indeed improve the capability of $\mu$VulDeePecker in multiclass vulnerability detection.
(iii) It is possible to further improve the effectiveness of $\mu$VulDeePecker by using it together with other deep learning based detection systems, which may be able to accommodate different kinds of information useful for multiclass vulnerability detection.

\subsection{$\mu$VulDeePecker vs. VulDeePecker}

We name our multiclass vulnerability detection system, namely $\mu$VulDeePecker, after the system known as VulDeePecker \cite{li2018vuldeepecker}, which is the first deep learning-based {\em binary} vulnerability detection system (i.e., only able to tell whether a piece of code is vulnerable or not, but not the type of a vulnerability in question). The reason we are so named is that our system is inspired by VulDeePecker; indeed, we refine the concept of {\em code gadget} introduced in \cite{li2018vuldeepecker}, which only captures data dependence, by additionally accommodating control dependence, which leads to higher effectiveness in multiclass vulnerability detection.
However, we stress that $\mu$VulDeePecker is {\em not} a simple incremental work over VulDeePecker \cite{li2018vuldeepecker}. 

The reasons are as follows.
First, our experiments show that it is not effective to extend VulDeePecker for multiclass vulnerability detection. For showing this, we consider two variants of VulDeePecker.
One variant, called VulDeePecker+, directly modifies the neural network architecture of VulDeePecker. The other variant is to train a VulDeePecker-based classifier for each {\em type} of vulnerabilities and each trained model is applied every sample for vulnerability detection. The latter variant has significant weaknesses (e.g., not scalable because there are many types of vulnerabilities; not effective when the samples of one vulnerability type are small), but is investigated for comparison purposes.
Second, we use new methods to prepare a dataset from scratch to evaluate the effectiveness of $\mu$VulDeePecker since the dataset published by the authors of VulDeePecker is not sufficient for our purposes. This is because (i) their dataset does not contain information on vulnerability types, (ii) their dataset only accommodates data-dependence but not control-dependence, and
(iii) their dataset loses some statements in the code gadgets, while these statements may contribute to pinpoint vulnerability types (as elaborated later).
Third, a novel concept of code attention and its extraction method are proposed in $\mu$VulDeePecker to help pinpoint vulnerability types.
Fourth, the neural network architecture underlying $\mu$VulDeePecker is more involved than the use of a standard BLSTM in VulDeePecker. The innovation in $\mu$VulDeePecker architecture lies in the fusion of different kinds of features; to the best of our knowledge, we are the first to use the fusion idea in the context of vulnerability detection.

\subsection{Paper Organization}
Section \ref{sec:definitions} discusses the basic ideas underlying $\mu$VulDee-Pecker and the terminology used in the paper. Section \ref{sec:system architecture} explains the detailed design of the system. Section \ref{sec:Implementation} introduces the implementation of the system. Section \ref{sec:Experiments and Analysis} is the systematic experiment and analysis of the results. Section \ref{sec:related work} is the related work. Section \ref{sec:limitation} discusses the limitation  and makes an explanation of the future work. Section \ref{sec:con} concludes the paper.

\section{Basic Ideas and Terminology}
\label{sec:definitions}

\input{definition}

\section{Design of $\mu$VulDeePecker}\label{sec:system architecture}
\input{design}

\section{Implementation and Evaluation Metrics}\label{sec:Implementation}
\input{implementation}

\section{Experiments and Results}
\label{sec:Experiments and Analysis}
\input{experiment}

\section{Related Work}\label{sec:related work}
Vulnerability detection in the source code is a fundamental problem of software security. Many scientists have conducted research on this issue and many good methods have been proposed. We divide prior studies on source-code vulnerability detection into three approaches: rule-based vs. similarity-based vs. pattern-based.

\subsection{Rule-Based Vulnerability Detection}
In this approach, vulnerability detection is based on some rules that are often defined by human experts. Open-source systems under this approach include
Flawfinder \cite{flawfinder}, Cvechecker \cite{cvechecker}, Cppcheck \cite{cppchecker}, FindBugs \cite{finderbugs}, and Splint \cite{splint}.
These tools use simple rules to characterize vulnerabilities and have limited successes.
Commercial systems under this approach include Checkmarx \cite{cx}, Coverity \cite{coverity}, IBM Security AppScan Source \cite{IBM}, and CodeSonar \cite{codesonar}. These commercial tools are more capable than the open-source tools mentioned above. The third category is the academic research methods. These academic research results are generally targeted at certain aspects of vulnerability detection and propose better vulnerability rules \cite{thome2017search, hackett2006modular, DBLP:journals/chinaf/FangLZWWW17}. This approach largely relies on human experts to define rules. In contrast, $\mu$VulDeePecker aims to automate the vulnerability detection process as much as possible, especially without relying on human experts to define rules with respect to every vulnerability.

\subsection{Similarity-Based Vulnerability Detection}
Since code-cloning is widely observed in practice, this approach aims to detect vulnerabilities that are often caused by code-cloning. The basic idea is that when a piece of code is cloned, the vulnerability in it is automatically replicated; this explains why detecting a piece of code that similar to a piece of vulnerable code could detect vulnerabilities. In this approach, similarity can be measured using {\em tokens} \cite{kamiya2002ccfinder, sajnani2016sourcerercc}, {\em strings} \cite{jang2012redebug, kim2017vuddy}, {\em trees} \cite{jiang2007deckard, pham2010detection}, {\em graphs} \cite{li2012cbcd}, or their hybrids \cite{li2016vulpecker}. This approach cannot detect vulnerabilities that are not caused by code-cloning \cite{li2018vuldeepecker}. $\mu$VulDeePecker does not follow this approach.

\subsection{Pattern-Based Vulnerability Detection}
Rather than relying on human experts for defining vulnerability rules, this approach is intended to define features and use machine learning techniques to {\em automatically} learn vulnerability patterns \cite{yamaguchi2011vulnerability, yamaguchi2012generalized, yamaguchi2013chucky, walden2014predicting, yamaguchi2014modeling, perl2015vccfinder, yamaguchi2015automatic}.
These studies still rely on human experts to define features to characterize vulnerabilities. A recent development is to exploit deep learning, which has a great potential in reducing the burden on human experts for defining features. VulDeePecker \cite{li2018vuldeepecker} is the first system using deep learning to detect vulnerabilities at slice-level, while noting that there are also related studies on using deep learning for vulnerability discovery at the function level \cite{lin2017poster, DBLP:journals/tii/LinZLPXVM18}, defect prediction \cite{wang2016automatically} and related tasks \cite{dam2018deep}. Above systems are binary classifiers by telling whether a piece of code is vulnerable or not.

The present study follows this approach and more specifically extends VulDeePecker to detect multiclass vulnerabilities. As discussed above, the extension is based on enhanced concept of code gadget, novel concept of code attention and a novel neural network architecture because a straightforward extension does not lead to good accuracy.

\section{Limitations}\label{sec:limitation}

The present study has limitations. First, $\mu$VulDeePecker can detect vulnerability types, but cannot pin down the precise location of a vulnerability at a granularity finer than code gadget. Although the granularity of code gadget, which consists of a number of statements, is substantially finer than the widely used granularity of programs, files or functions, it is an interesting future work to precisely pin down the location (e.g., exactly the vulnerable statements but nothing more) of a vulnerability. This is important because it will alleviate human analysts in pinning down locations of vulnerabilities. Second, the current design and implementation of $\mu$VulDeePecker are geared towards programs written in C/C++. future research needs to consider programs written in other programming languages. Third, the current design and implementation of $\mu$VulDeePecker focus on vulnerabilities that are related to library/API function calls. Future research needs to consider vulnerabilities that are not associated to library/API functions. 

\section{Conclusion}\label{sec:con}

We have presented $\mu$VulDeePecker, which is the first deep learning-based multiclass vulnerability detection system. Its capability in pinning down the type of vulnerability in a code gadget (i.e., a number of statements) helps human analysts in recognizing vulnerabilities. The multiclass detection capability largely comes from the use of code attentions. Systematic experiments show that $\mu$VulDeePecker is effective, that accommodating control-dependence can lead to higher detection capabilities, and that $\mu$VulDeePecker can be used together with other systems (e.g., VulDeePecker+) to capture more useful information for multiclass vulnerability detection. The limitations discussed above are interesting open problems for future investigations.

%% file: definition.tex

In this section we describe the key idea underlying $\mu$VulDeePecker and rigorously define the terminology used in the present paper.

\subsection{Basic Ideas}

In order to detect the specific types of vulnerabilities, we propose using the concept of {\em code attention}, which is inspired by the notion of {\em region attention} in the image processing \cite{fu2017look}. The notion of region attention in image processing was introduced to capture the insight that some regions in an image provide more discriminative information for accurate classification of the image. For example, when human eyes observe an image containing a bird, the bird's eyes, mouth, and coat color provide more information than other regions for identifying the type of the bird. Similarly, code attention is composed of multiple program statements in a piece of code and would provide more information for classifying the type of a vulnerability.
For example, when a piece of code is detected as a vulnerability caused by illegal library/API function calls, argument definition statements in library/API function calls, control statements, and statements containing library/API function calls may provide more information for classifying the types of vulnerabilities.

The concept of code attention refines the concept of {\em code gadget}, which is the key idea underlying the first deep learning-based vulnerability detection system known as VulDeePecker \cite{li2018vuldeepecker}. A code gadget, as defined in \cite{li2018vuldeepecker}, is a number of (not necessarily consecutive) statements that accommodate the {\em data-dependence} relation.
A code gadget captures some degree of syntax and semantic information of vulnerabilities, explaining why it helps detect whether a piece of code is vulnerable or not.
As we will discuss later, we further use a refined version of the notion of code gadget, which captures not only the data-dependence relation but also the {\em control-dependence} relation.

We propose using both the aforementioned refined notion of code gadget and the aforementioned code attention for training the multiclass vulnerability detection model.
Intuitively, the refined notion of {\em code gadget} captures more ``global'' semantics information, which is conveyed by the control-dependence and data-dependence relations between the statements in a program, and thus would help achieve a higher capability in detecting whether a piece of code is vulnerable or not. The notion of {\em code attention} captures more ``localized'' information within a statement (e.g., arguments in a specific library/API function call) and would help recognize the specific type of a vulnerability. More specifically, code attention enforces data source checking, critical path checking (condition statements), and dangerous function usage checking; these factors are closely related to the types of vulnerabilities.
In order to attain a full-fledged solution, we propose fusing these two kinds of information into a more comprehensive feature representation.

\subsection{Definitions}

In order to make our description precise, we use the following definitions.

\begin{defn}[program, statement and token]
A program $P$ is an ordered set of statements, denoted by $P=\{p_1, p_2, \ldots, p_\varepsilon\}$, where $p_i$ ($1 \le i \le \varepsilon$) is a statement. A statement $p_i$ is an ordered set of tokens, denoted by $p_i=\{t_{i,1},t_{i,2},\ldots,t_{i,w}\}$, where token $t_{i,j}$ ($1 \le j \le w$) can be a variable identifier, function identifier, constant, keyword or operator and so on.
\end{defn}

\begin{defn}[data-dependence \cite{tip1994survey}]
Given a program $P=\{p_1, p_2, \ldots, p_\varepsilon\}$, a statement $p_i\in P$, a variable identifier token $t_{i,j}$ belonging to $p_i$, statement $p_u$ ($1 \le u \le \varepsilon$) is said to be {\em data-dependent} on token $t_{i,j}$ if $t_{i,j}$ is used in $p_u$.
\end{defn}

\begin{defn}[control-dependence \cite{tip1994survey}]
Consider two statements $p_i, p_j$ ($i\neq j$) in a program $P=\{p_1, p_2, \ldots, p_\varepsilon\}$. If the execution outcome of $p_i$ affects whether $p_j$ will be executed or not, $p_j$ is said to be control-dependent on $p_i$.
\end{defn}


\begin{defn}[code gadget \cite{li2018vuldeepecker}]
Consider a program $P=\{p_1, p_2, \ldots, p_\varepsilon\}$, a library/API function call denoted by $f_{v_i}$, and a statement $p_v$ ($1 \le v \le \varepsilon$) containing the function call $f_{v_i}$. Assume that $p_v$ contains a total of $m$ function calls, then $1 \le v_i \le m$. Denote the arguments of the function call $f_{v_i}$ by a set $A_{v_i}=\{t_{v_i,x_1},t_{v_i,x_2},\ldots,t_{v_i,x_\eta}\}$.
A {\em code gadget} corresponding to function call $f_{v_i}$ in statement $p_v$, denoted by $s_{v_i}$, is an ordered set of statements in $P$, each of which is either recursively {\em data-dependent} on one or multiple arguments in $A_{v_i}$ or {\em control-dependent} on $p_v$ in a recursive manner.
\end{defn}


\begin{defn}[code attention]
Consider a program $P=\{p_1, p_2, \ldots, p_\varepsilon\}$, a library/API function call $f_{v_i}$, a statement $p_v$ ($1 \le v \le \varepsilon$) containing the function call $f_{v_i}$, and a code gadget $s_{v_i}$ corresponding to $f_{v_i}$ in $p_v$. Let $R =\{r_k\}_{1 \leq k \leq h}$ be a set of rules describing vulnerability syntax characteristics, which are vulnerability-specific and thus elaborated later. A {\em code attention} with respect to code gadget $s_{v_i}$, denoted by $c_{v_i}$, is a subset of the statements in code gadget $s_{v_i}$ that match the syntax characteristics specified by some $r_k \in R$.
\end{defn}

%% file: design.tex
Our objective is to design a multiclass vulnerability detection system.
Let $\{0,1, 2, \ldots, m\}$ denote a set of vulnerability types, where type-0 means ``not vulnerable'' and type-$i$ ($1\leq i \leq m$) corresponds to a Common Weakness Enumeration IDentifier or CWE-ID (which is the outcome of a community effort at categorizing vulnerabilities \cite{cwe}).
As the first study on multiclass vulnerability detection, we focus on vulnerabilities related to library/API function calls in C/C++ programs, while leaving the extension to accommodating other vulnerabilities to future work. 

\begin{figure*}[!htbp]
	\centering
	\includegraphics[width=\textwidth]{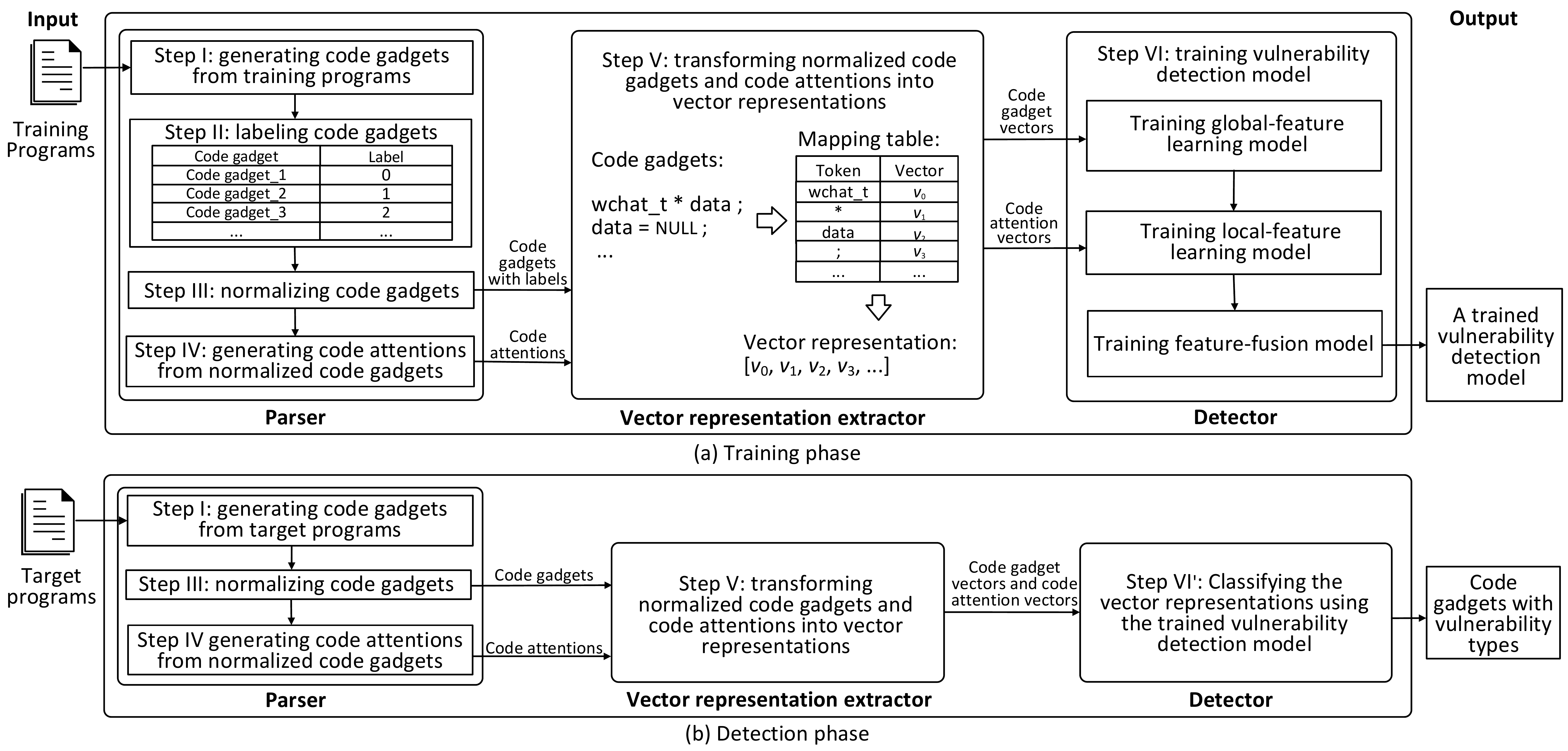}
	\caption{Overview of $\mu$VulDeePecker's three modules (i.e., parser, vector representation extractor, and detector) used in the training and detection phases.}
	\label{fig:overview}
\end{figure*}

\begin{figure*}[!htbp]
	\centering
	\includegraphics[width=\textwidth]{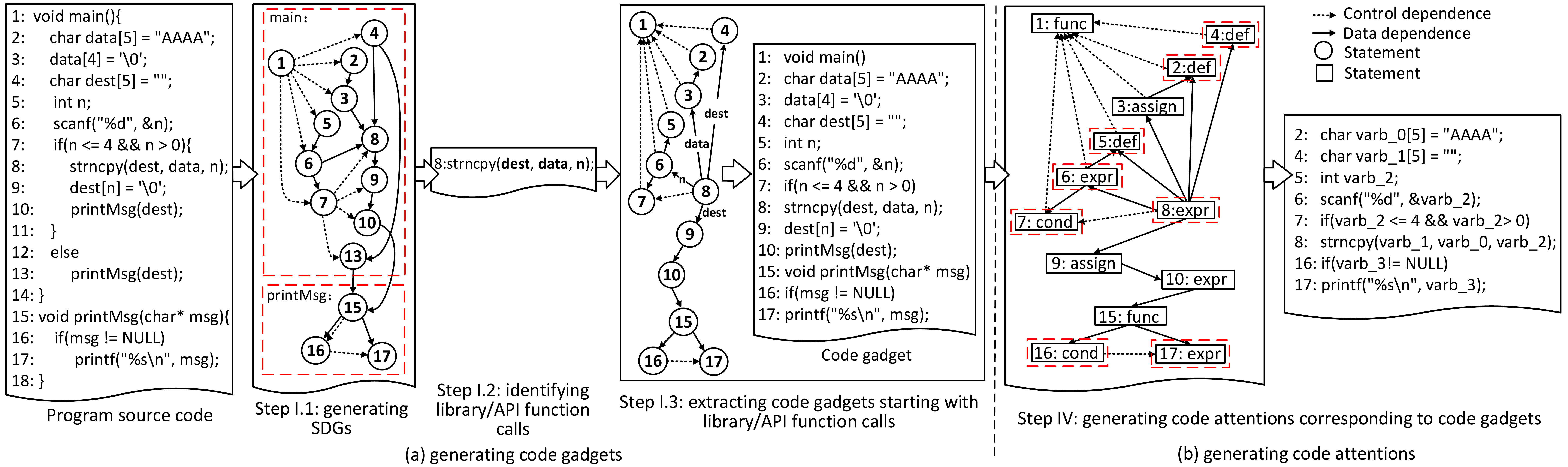}
	\caption{(a) An example showing the generation of code gadget corresponding to function call $strncpy$ in Line 8 of the program. 
(b) An example showing the code attention derived from the code gadget, namely statements matching vulnerability syntax characteristics (highlighted in red-colored boxes).}
	\label{fig:exsa}
\end{figure*}

We design $\mu$VulDeePecker in three modules: (i) parser, which parses programs into code gadgets and code attentions; (ii) vector representation extractor, which generates vector representation of code gadgets and code attentions; (iii) detector, which learns a vulnerability detection model and uses this model for multiclass vulnerability detection. As highlighted in Fig \ref{fig:overview} and elaborated below, both the training phase and the detection phase contain these three modules, except that Step II (i.e., ground-truth labeling in the training phase) is not relevant to the detection phase because labels are pursued as the output of the detection phase and that Step VI is about training a detector and Step VI' is about using the trained detector. 

\subsection{Training Phase}\label{sec:the workflow of training phase}

As highlighted in Fig \ref{fig:overview}$(a)$, the parser is divided into Steps I-IV, the vector representation extractor corresponds to Step V, and the detector corresponds to Step VI. These steps are elaborated below.

\textbf{Step I: } this step generates code gadgets from training programs. Since this step is quite involved, we use the example shown in Fig \ref{fig:exsa}$(a)$ to further illustrate ideas. It can be divided into 3 sub-steps.

\textbf{Step I.1:} this step generates a System Dependency Graph (SDGs) \cite{sinha1999system} for each training program. A SDG is derived from a set of Program Dependence Graphs (PDGs) \cite{ferrante1987program}, which represent data-dependence and control-dependence relations. A PDG is a directed graph, wherein a node (or vertex) represents a statement or control predicate and an arc (or directed edge) represents a data- or control-dependence relation between two nodes. There are standard algorithms for generating PDGs for functions (see, for example, \cite{ferrante1987program}). A SDG can be derived from PDGs via the caller-callee relation. Specific to the purpose of the present paper, we propose associate to each node with attributes, including the statement text corresponding to the node and the type of the statement (e.g. ``expr'', ``def'', ``assign''). The second column in Fig \ref{fig:exsa}$(a)$ shows the SDG derived from the PDGs of functions $main$ and $printMsg$ in the example program, where each node is only highlighted with the corresponding statement (represented by the Line number) for succinctness.

\textbf{Step I.2:} this step identifies library/API function calls in programs of interest.
It extracts library/API function calls by matching statements associated to nodes to known library/API function calls. For example, the third column in Fig \ref{fig:exsa}$(a)$ contains the library/API function call $strncpy$ in Line 8 of the example program.
	\begin{algorithm}[!htbp]
		\algrenewcommand\algorithmicrequire{\textbf{Input:}}
		\algrenewcommand\algorithmicensure{\textbf{Output:}}
		\caption{Extracting code gadgets (Step I.3)}
		\label{alg:bi-algorithm}
		
		\begin{algorithmic}[1]
			\Require A SDG vertex $n$ corresponding to statement $p_v$ which contains library/API function call $f_{v_i}$; the arguments set $A_{v_i}$ of function call $f_{v_i}$.
			\Ensure A code gadget $s_{v_i}$ corresponding to $f_{v_i}$ in $p_v$.\vspace{0ex}	
			\newline
			
			\Function{$get\_for\_slice$}{$t_{v_i,x_\delta}, n, sf_{v_i}$}
			
			\State $N_{su} \leftarrow$ the successor vertices set of $n$
			\For{ each $n_{su} \in N_{su}$}
			\If {$n_{su}$ is data dependent on $t_{v_i,x_\delta}$ or control dependent on $n$}
			\State $sf_{v_i} \leftarrow sf_{v_i} \cup \{n_{su}\}$
			\State $n \leftarrow n_{su}$
			\State \Call {$get\_for\_slice$}{$t_{v_i,x_\delta}, n, sf_{v_i}$}
			\EndIf
			\EndFor
			
			\EndFunction
			
			\Function{$get\_back\_slice$}{$t_{v_i,x_\delta}, n, sb_{v_i}$}
			
			\State $N_{pre} \leftarrow$ the predecessor vertices set of $n$
			\For{ each $n_{pre} \in N_{pre}$}
			\If {$n_{pre}$ is data dependent on $t_{v_i,x_\delta}$ or control dependent on $n$}
			\State $sb_{v_i} \leftarrow sb_{v_i} \cup \{n_{pre}\}$
			\State $n \leftarrow n_{pre}$
			\State \Call {$get\_back\_slice$}{$t_{v_i,x_\delta}, n, sb_{v_i}$}
			\EndIf
			\EndFor
			
			\EndFunction

			\Function{$main$}{$A_{v_i}, n$}
			\State $s_{v_i} \leftarrow  \emptyset$
			\State $sf_{v_i} \leftarrow  \emptyset$
			\State $sb_{v_i} \leftarrow  \emptyset$
			
			\For {each $t_{v_i,x_\delta} \in A_{v_i}$}
			\State \Call{$get\_for\_slice$}{$t_{v_i,x_\delta}, n, sf_{v_i}$}
			\State \Call{$get\_back\_slice$}{$t_{v_i,x_\delta}, n, sb_{v_i}$}
			\EndFor
			
			\State $s_{v_i} \leftarrow sf_{v_i}\cup sb_{v_i}$
			\State \Return $s_{v_i}$
			\EndFunction
		\end{algorithmic}
	\end{algorithm}

\textbf{Step I.3:} this step extracts code gadgets using a method that is based on \cite{tip1994survey} but is different from the method used by VulDeePecker \cite{li2018vuldeepecker}. We note that VulDeePecker (i) only considers data-dependence, (ii) divides functions into forward functions (which receive external inputs) and backward functions (which do not receive external inputs), and (iii) extracts forward (backward) slices from forward (backward) function calls. The forward slice is obtained by slicing the successors of the forward function call. Conversely, the backward slice is obtained by slicing the precursors of the backward function call. In contrast, we accommodate both data-dependence and control-dependence when extracting code gadgets in a bi-direction manner (i.e., considering both forward and backward slices). Our approach to extracting code gadgets is advantageous to the approach used by VulDeePecker because (i) our approach accommodates both data-dependence and control-dependence in code gadgets, rather than data-dependence only, and (ii) our approach considers {\em all} of the forward and backward slices rather than extracting forward or backward slices by selecting some forward or backward functions, which is the case in VulDeePecker \cite{li2018vuldeepecker}.
	
In order to see the preceding (ii), we consider the library functions $scanf$ and $malloc$ as examples, which represent the forward function of the input and the backward function of the non-input. The function $scanf$ would be defined as a forward function and therefore lead to a forward slice in VulDeePecker. However, the definition statements corresponding to the arguments of $scanf$ can only be contained in the backward slice of $scanf$, which leads to these statements miss. These statements missed by VulDeePecker would be important because they provide the information about the types of these arguments and their memory sizes, which can be leveraged as a reference when examining whether these arguments have improper operations or not. For another example of library function $malloc$, which is defined as a backward function in \cite{li2018vuldeepecker}, resulting in generating a backward slice. It also causes that the operation statements of the memory allocated by $malloc$ miss because these statements only can be sliced in the forward slice of $malloc$. But these operations are important as they are the basis for judging whether memory leak or buffer overflow has occurred. Therefore, we believe that only considering forward slice or backward slice will cause information loss, which may be useful in vulnerability detection. That is why we consider both forward and backward slices in the present paper.
	
Algorithm \ref{alg:bi-algorithm} elaborates Step I.3. Consider a library/API function call $f_{v_i}$ made in a statement $p_v$, which is represented as a node $n$ in the SDG in question. For each argument $t_{v_i,x_\eta}$ of $f_{v_i}$, Algorithm \ref{alg:bi-algorithm} does the following: (i) using function $get\_for\_slice$ (Lines 1-10 in Algorithm \ref{alg:bi-algorithm}) to identify successors of node $n$ in the SDG, namely the nodes that are data-dependent upon $t_{v_i,x_\delta}$ or control-dependent upon $n$ (either directly or indirectly), and then generates a forward slice $sf_{v_i}$; (ii) using function $get\_back\_slice$ (Lines 11-20 in Algorithm \ref{alg:bi-algorithm}) to identify predecessors of node $n$ recursively, upon which $t_{v_i,x_\delta}$ is data-dependent or $n$ is control-dependent (either directly or indirectly), and then generates a backward slice $sb_{v_i}$. Finally, Algorithm \ref{alg:bi-algorithm} merges slices $sf_{v_i}$ and $sb_{v_i}$ (Line 29 in Algorithm \ref{alg:bi-algorithm}) to obtain a code gadget $s_{v_i}$ corresponding to function call $f_{v_i}$. 
The last column in Fig \ref{fig:exsa}$(a)$ shows the example of extracting the code gadget according to arguments $data$,  $dest$ and $n$ of function call $strncpy$.

\textbf{Step II:} this step labels each code gadget as ``0'' (not vulnerable) or $i$ (type-$i$ vulnerability), where $1\leq i \leq m$. 
These labels are the ground truth for training a multiclass vulnerability detection model.

\begin{figure}[!htbp]
	\centering
	\includegraphics[width=.48\textwidth]{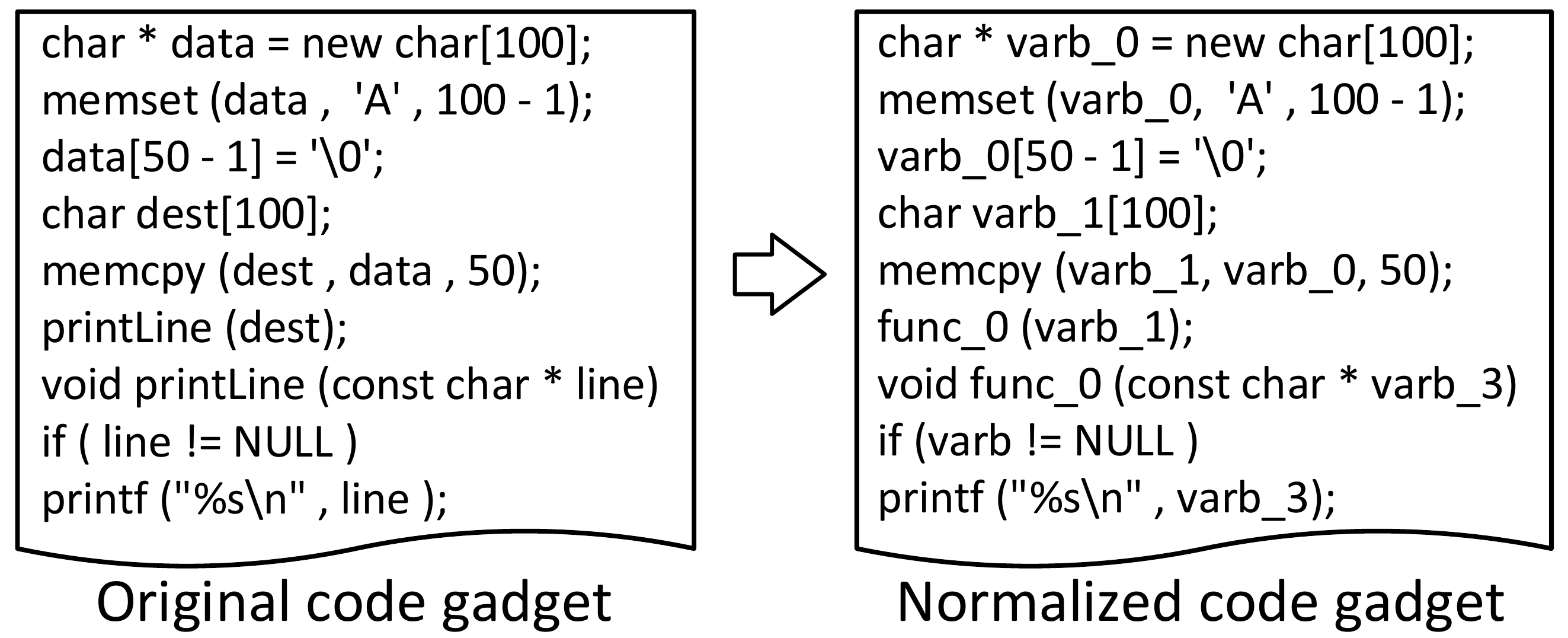}
	\caption{Example of illustrating code gadget normalization.}	
	\label{fig:module2}
\end{figure}

\textbf{Step III:} we observe that different programmers have different coding habits (e.g., format, variables, and function naming), which may affect the capability of multiclass vulnerability detection models. This motivates us to normalize code gadgets by (i) mapping same variables and functions to same values and (ii) renaming different variables and functions according the order of their appearance (i.e., ``varb\_0'', ``varb\_1'', $\ldots$; ``func\_0'', ``func\_1'', $\ldots$). However, reserved words, library/API function names, and constant are not mapped because they are standard.
Fig \ref{fig:module2} gives an example of code gadget normalization. 

\textbf{Step IV:} this step parses the normalized code gadgets to generate code attentions according to vulnerability syntax characteristics, which would contribute to the identification of vulnerability types. For the vulnerabilities caused by improper library/API function calls, we observe that the arguments and usage of the library/API function calls affect the occurrence of vulnerabilities. Taking into consideration data source checking, data sanitized operation checking, and library/API function calls usage plausibility checking, we propose utilizing the following three syntax characteristics:
(i) definition statements of arguments in library/API function calls, which would provide information useful to identify improper use of library/API function calls (e.g., type and memory size of an argument) and distinguish whether the vulnerability caused by data source;
(ii) control statements, which would help determine whether a program has conducted proper bounds-checking and security screening before executing the target library/API function calls; 
(iii) statements containing library/API function calls, which would directly help recognize vulnerability types.
Corresponding to these vulnerability syntax characteristics, we define rules as follows:
\begin{itemize}
\item Rule $r_1$: if the attribute of a statement $p_i$ in code gadget $s_{v_i}$ is a definition statement and the variables defined in $p_i$ match the arguments in the library/API function call, then $p_i$ is a statement in code attention $c_{v_i}$. For example, statements 2, 4  and 5 in Fig \ref{fig:exsa}$(b)$ are definition statements of arguments $varb\_0$, $varb\_1$ and $varb\_2$, which are used in function call $strncpy$.
	
\item Rule $r_2$: if the attribute of statement $p_i$ in code gadget $s_{v_i}$ is a control statement, then $p_i$ is a statement in code attention $c_{v_i}$. For example, statements 7 and 16 in Fig \ref{fig:exsa}$(b)$ are such statements.
	
\item Rule $r_3$: if the attribute of statement $p_i$ in code gadget $s_{v_i}$ contains library/API function calls, then $p_i$ is a statement in code attention $c_{v_i}$. For example, statements 6, 8, and 17 in Fig \ref{fig:exsa}$(b)$ are such statements.
\end{itemize}

	\begin{algorithm}[!htbp]
		\algrenewcommand\algorithmicrequire{\textbf{Input:}}
		\algrenewcommand\algorithmicensure{\textbf{Output:}}
		\caption{Extracting code attention (Step IV)}
		\label{alg:2}
		\begin{algorithmic}[1]
			\Require Normalized code gadget $s_{v_i}$; a set $R =\{r_k\}_{1 \leq k \leq h}$ of rules to match syntax characteristics of vulnerabilities.
			\Ensure Code attention $c_{v_i}$ in code gadget $s_{v_i}$.
			
			\State $c_{v_i} \leftarrow \emptyset$
			\For{each statement $p_i \in s_{v_i}$}
			\State Parse $p_i$ into an ordered set of tokens $\{t_{i,1},t_{i,2},\ldots,t_{i,w}\}$ via a lexical analysis,
             obtaining attribute $attr$ for statement $p_i$
			\For{each $r_k \in R$}
			\For{each token $t_{i,j} \in \{t_{i,1},t_{i,2},\ldots,t_{i,w}\}$}

			\If{$attr$ and $t_{i,j}$ match $r_k$}
			\State $c_{v_i} \leftarrow c_{v_i} \cup \{p_i\}$
			
			\EndIf
			\EndFor
			\EndFor

			\EndFor
			
			\State \textbf{return }$c_{v_i}$
			
		\end{algorithmic}
	\end{algorithm}

Using the aforementioned vulnerability syntax characteristics, Algorithm \ref{alg:2} extracts a code attention from a normalized code gadget. 
The algorithm can be understood as follows. 
For each statement $p_i$ in $s_{v_i}$, Algorithm \ref{alg:2} parses statement $p_i$ into an ordered set of tokens by a lexical analysis, infers attribute $attr$ of statement $p_i$ (e.g., ``def'', ``cond'') by analyzing the context structure of these tokens, and uses each token and $attr$ to match the rules in a given rule set $R =\{r_k\}_{1 \leq k \leq h}$ to generate code attentions.
Currently, we only use a rule set $R=\{r_1,r_2,r_3\}$ of the three rules mentioned above.

\textbf{Step V:} this step converts normalized code gadgets and code attentions respectively to fixed-length code gadget vectors and code attention vectors. The conversion is attained by a simple {\em word embedding} \cite{mikolov2013efficient}.
These vectors are the input for training neural networks.

\begin{figure}[!htbp]
	\centering	
	\includegraphics[width=.48\textwidth]{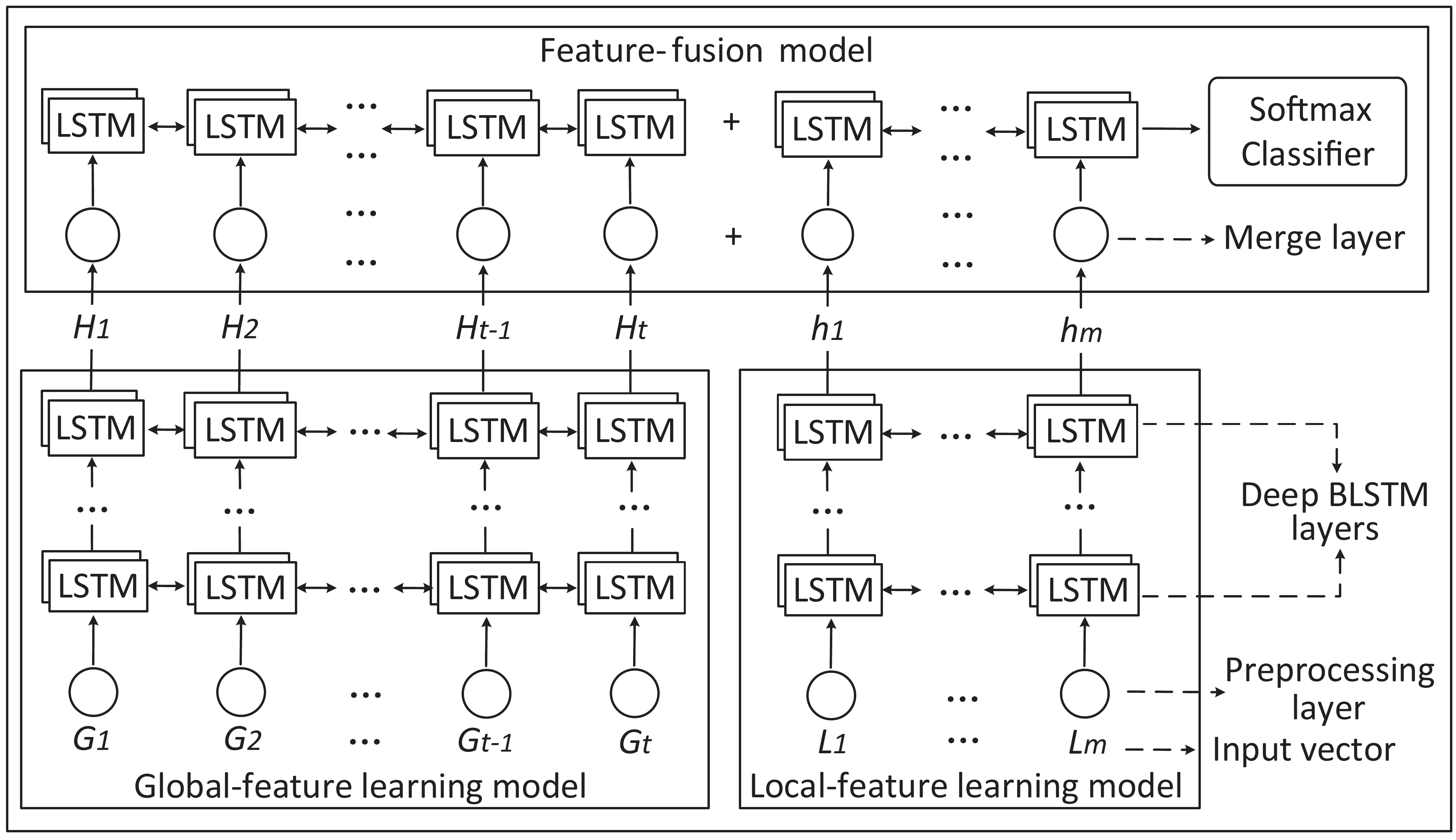}
	\caption{The $\mu$VulDeePecker neural network architecture.}
	\label{fig:mode_arch}
\end{figure}

\textbf{Step VI:} inspired by the multi-feature fusion method in image and video recognition \cite{ou2017adult}, we propose a new neural network architecture that uses 
the BLSTM as a building-block because BLSTM is shown to be capable of detecting vulnerabilities \cite{li2018vuldeepecker}.
The network, as highlighted in Fig \ref{fig:mode_arch}, deals with two kinds of features:
{\em global features}, which are denoted by $H_i$'s in Fig \ref{fig:mode_arch}, are learned from code gadgets, and accommodate some broader semantics about the relations between statements in a program; {\em local features}, which are denoted by $h_i$'s in Fig \ref{fig:mode_arch}, are learned from code attentions, and are specific to individual statements in a program (e.g., arguments in a specific library/API function call). 
Since these two kinds of features accommodate different kinds of information, we use a feature fusion to learn comprehensive features. As a result, the network consists of the following three BLSTM networks: global-feature learning model, local-feature learning model, and feature-fusion model.
For learning {\em global features} from code gadgets, the model uses a pre-processing layer and deep BLSTM layers, where the preprocessing layer mainly filters the $0$ vectors and the deep BLSTM layers actually learn global features. Each neuron is equivalent to a time step and receives a word in a code gadget.
For learning {\em local features} from code attentions, the model uses the same network (for learning global features) but at a smaller scale (because vectors representing code attentions are shorter than vectors representing code gadgets).
For fusing global and local features, the model uses a merge layer, a BLSTM layer and a softmax classifier, where the merge layer connects the learned global and local features and the BLSTM layer adjusts the spliced features. Finally, the fused features are loaded into the classifier.

Summarizing the preceding discussion, we first train a global-feature learning model as well as a local-feature learning model and then train the feature-fusion model.
The resulting system is used for multiclass vulnerability detection. 
Details of the training strategy are elaborated in Section \ref{sec:imple_train} with respect to a specific training phase.

\subsection{Detection Phase}
\label{sec:the workflow of detecting phase}
As highlighted in Fig \ref{fig:overview}$(b)$, this phase contains the following steps.

\textbf{Step I:} this step generates code gadgets from target programs for vulnerability detection (and is similar to Step I in the training phase).

\textbf{Step III:} this step normalizes code gadgets (and is similar to Step III in the training phase).	

\textbf{Step IV:} this step extracts code attentions from code gadgets (similar to Step IV in the training phase).
		
\textbf{Step V:} this step transforms the normalized code gadgets and code attentions to vector representations (and is similar to step V in the training phase).

\textbf{Step V':} this step classifies vector representations using the learned $\mu$VulDeePecker model, outputting the vulnerability types of code gadgets (recalling that type-0 means the code gadgets is not vulnerable).

\subsection{Generalization Analysis}
$\mu$VulDeePecker can only detect vulnerabilities caused by C/C++ library/API function calls currently, but the techniques used are extensible (e.g., extracting code gadgets and code attentions). $\mu$VulDeePecker can be adopted/adapted to accommodate other programming languages and other types of vulnerabilities, but the generalization should adhere to the following premise:
\begin{itemize}
	\item The software source code to be generalized must be able to be parsed into SDG structures.
	\item The type of vulnerability to be generalized must have an interesting entity. For example, the entity of interest in this paper is C/C++ library/API function calls.
	\item The type of vulnerability to be generalized must exhibit the following 3 perspectives:  data source checking, data sanitized operation checking, and interested entity usage plausibility checking.
\end{itemize}

If the target language and target vulnerability type to be generalized satisfy the above premises, $\mu$VulDeePecker can be extended, which is also the future work of us.

%% file: implementation.tex
In this section, we describe the preparation of the dataset for training and testing and the implementation of $\mu$VulDeePecker.

\subsection{Dataset Preparation}
\label{sec:data}

Our sources of vulnerabilities are Software Assurance Reference Dataset (SARD) \cite{sard} and National Vulnerability Database (NVD) \cite{nvd}.

SARD provides a large set of production, synthetic, and academic programs (a.k.a. test cases) with known vulnerabilities. A program is labeled as ``good'' (i.e., not vulnerable), ``bad''(i.e., vulnerable), or ``mixed'' (i.e., both vulnerable and patched versions are available).
For a vulnerable program, SARD provides the statements containing the vulnerability and the vulnerability type in CWE-ID (Common Weakness Enumeration IDentifier) \cite{cwe}.
NVD contains a large number of vulnerabilities in production software and
provides the software products affected, the vulnerable versions, the CWE-IDs, and the patch files.
In total, we collect 33,409 programs written in C/C++ corresponding to 116 CWE-IDs, including (i) 57 ``good'' programs, 925 ``bad'' programs, and 3,2104 ``mixed'' programs collected from SARD and (ii) 323 vulnerable programs collected from NVD.
Since CWE-IDs are hierarchic (e.g., CWE-121 contains CWE-787 and CWE-788 as sub-types),
we aggregate CWE-IDs to the third level of the CWE-ID tree \cite{cwe_relation} (i.e., the {\em research concept view}) and use these third-level CWE-IDs as vulnerability types.
For example, CWE-121 is a sub-type of CWE-119, which is at the third level of the CWE-ID tree, and as such we use CWE-119 as the vulnerability type for a CWE-121 vulnerability as well.
This leads to 40 different vulnerability types (i.e., $m=40$ plus the non-vulnerability type), which are listed in Table \ref{t:labels_type}. We observe that most vulnerabilities do not contain vulnerability sub-types, but some of them do (e.g., Type-7 or CWE-415 is a sub-type contained in three types CWE-119, CWE-666, and CWE-573); this is caused by the CWE-ID tree structure mentioned above.
We randomly respectively select 80\% SARD programs and 80\% NVD programs as the training set and the rest as the testing set for experiments.

\begin{table}[!htbp]
	\small
	\centering
	\caption{The specific vulnerability types and their corresponding labels of the 40 types of vulnerabilities in this paper.}
	\begin{tabular}{m{0.6cm}|m{2.8cm}|m{0.6cm}|m{2.8cm}}
		\toprule
		Labels & Vulnerability Types &Labels & Vulnerability Types\\
		\hline
		1 & CWE-404 & 21 & CWE-170\\
		\hline
		2 & CWE-476 & 22 & CWE-676\\
		\hline
		3 & CWE-119 & 23 & CWE-187\\
		\hline
		4 & CWE-706 & 24 & CWE-138\\
		\hline
		5 & CWE-670 & 25 & CWE-369\\
		\hline
		6 & CWE-673 & 26 & CWE-662, CWE-573\\
		\hline
		7 & CWE-119, CWE-666, CWE-573 &27 & CWE-834\\
		\hline
		8 & CWE-573 & 28 & CWE-400, CWE-665\\
		\hline
		9 & CWE-668 & 29 & CWE-400, CWE-404\\
		\hline
		10 & CWE-400, CWE-665, CWE-020 & 30 & CWE-221\\
		\hline
		11 & CWE-662 & 31 & CWE-754\\
		\hline
		12 & CWE-400 & 32 & CWE-311\\
		\hline
		13 & CWE-665 & 33 & CWE-404, CWE-668\\
		\hline
		14 & CWE-020 & 34 & CWE-506\\
		\hline
		15 & CWE-074 & 35 & CWE-758\\
		\hline
		16 & CWE-362 & 36 & CWE-666\\
		\hline
		17 & CWE-191 & 37 & CWE-467\\
		\hline
		18 & CWE-190 & 38 & CWE-327\\
		\hline
		19 & CWE-610 & 39 & CWE-666, CWE-573\\
		\hline
		20 & CWE-704 & 40 & CWE-469\\
		\bottomrule
	\end{tabular}
	\label{t:labels_type}
\end{table}

\subsection{Training Phase}
\label{sec:imple_train}

This phase is divided into the following six Steps.

\textbf{Step I:} for generating code gadgets from training programs, we use the open source C/C++ code analysis tool Joern \cite{joern} to construct PDGs for functions in each training program.
Then, we generate an SDG according to the caller-callee relation between these functions, and traverse each node in the SDG to identify the nodes that contain
any of the 811 C/C++ library/API function calls related to security which are publicly available in the rules of Checkmarx \cite{cx}.
Finally, we extract the arguments of each function call and use Algorithm \ref{alg:bi-algorithm} to generate code gadgets.

\textbf{Step II:} the vulnerable statements in a vulnerable program collected from SARD are readily available from the dataset.
The vulnerable statements in a vulnerable program collected from NVD are the statements that are deleted in the patch files.
If a code gadget contains one or more vulnerable statements, it is labeled as vulnerable with vulnerability type-$i$ where $1\leq i \leq 40$; otherwise, it is labeled as type-$0$ (i.e., non-vulnerable).

\begin{table*}[!htbp]
	\centering
	\small	
	\caption{The hyperparameter values of the global-feature learning model, local-feature learning model, and feature-fusion model.}
	\begin{tabular}{m{3.5cm}|m{4.4cm}|m{4.2cm}|m{3.8cm}}
		\toprule
		Hyper-parameters&Global-feature learning model &Local-feature learning model & Feature-fusion model\\
		\hline
		Optimizer &RMSprop &RMSprop &RMSprop \\
		\hline
		Loss function &categorical\_crossentropy
		 &categorical\_crossentropy
		  &categorical\_crossentropy\\
		  \hline
		Activation function&tanh &tanh &tanh \\
		\hline
		
		Learning rate &0.001 &0.001 &0.001 \\
		\hline
		Batch size & 64 & 64 & 64 \\
		\hline
		Dropout &0.5 &0.5 &0.5 \\
		\hline
		Layers &2 &2 &1 \\
		\hline
		Nodes in a hidden layer&300 &200 &500 \\
		\hline
		Epochs &60 &60 &10 \\
		\bottomrule
	\end{tabular}
	\label{t:param}	
\end{table*}

\textbf{Step III and Step IV:} for each code gadget, we write an automate lexical-analysis-based program to analyze each token (e.g., variables, operators, keywords) in statements. We mainly analyze the types and the context structures of them for identifying the variable and function names and rename them. Finally, we use Algorithm \ref{alg:2} on the normalized code gadgets to extract code attentions.

\textbf{Step V:} for normalized code gadgets, we use a lexical analysis to generate a corpus of program tokens. Then, we use the word-to-vector tool \cite{word2vec} to generate the vectors for representing those tokens. The word-to-vector model we use is skip-gram \cite{DBLP:conf/nips/MikolovSCCD13}, with window size 10 and vector dimension 50.
Since the vector input to the neural network is fixed-length,
we use $\tau_1$ and $\tau_2$ respectively for the length of code gadgets and the length of code attentions,
while noting that these hyperparameters are tuned in the training phase.
For code gadgets (code attentions) containing fewer words or shorter vectors than $\tau_1$ ($\tau_2$), we pad 0 vectors at their tail to make their vectors of length $\tau_1$ ($\tau_2$);
for code gadgets (code attentions) containing  more words or longer vectors than  $\tau_1$ ($\tau_2$), we cut their vectors at the tail of code gadgets (code attentions) to match length $\tau_1$ ($\tau_2$).

\textbf{Step VI:} recall that we use the BLSTM as a building-block for constructing $\mu$VulDeePecker neural network architecture.
Our implementation of the model uses Keras \cite{keras}.
We first train global-feature and local-feature learning models by tuning parameters --- ``optimizer'', ``learning rate'', ``batch size'', ``dropout'', ``layers'', ``the number of nodes in a hidden layer'', and ``epochs'' --- to achieve the optimal results via the method called {\em grid search}, which searches the optimal values of these hyperparameters by performing an exhaustive search. Next, we search for the best fusion method via the permutation and combination of global-feature and local-feature models, while fixing the hyperparameters. Finally, we tune the hyperparameters of the feature-fusion model and obtain the $\mu$VulDeePecker neural network.
We train the three networks separately so as to prevent the features learned from one network from being destroyed by the others. The specific hyperparameter values of each model obtained in training phase are summarized in Table \ref{t:param}.

\subsection{Detection Phase}
\label{sec:the workflow of detecting phase}

In the detection phase, we perform 5 steps for target programs, of which the first 4 steps are the same as Step I, III, IV, V in the training phase. In Step VI', we input the code gadget vectors and their corresponding code attention vectors to be tested into the trained $\mu$VulDeePecker model. The model outputs whether the code gadgets contain vulnerabilities and if so, outputs their classes.

Totally, we extract 181,641 code gadgets (145,353 in the training set and 36,288 in the testing set) whose average/median number of lines-of-code is 140/120, among which 175,415 are from SARD (including 42795 vulnerable and 132620 non-vulnerable) and 6,226 from NVD (including 324 vulnerable and 5902 non-vulnerable). To sum up, there are 138,522 non-vulnerable code gadgets and 43,119 vulnerable code gadgets covering 40 vulnerability types which
are listed in Table \ref{t:labels_type}. This labeled dataset, called Multiclass Vulnerability Dataset (MVD), is available at \url{https://github.com/muVulDeePecker/muVulDeePecker}.

\subsection{Evaluation Metrics}

For evaluating multiclass vulnerability detectors, we use the following metrics (see, for example, \cite{metric}).
Let $L$ be a set of vulnerability types (in this paper, $|L|=m=40$),
Let $TP_l$, $FP_l$, $FN_l$, $TN_l$ and $X_l$ respectively be the number of true positive samples, false positive samples, false negative samples, true negative samples, and the total number of samples with respect to vulnerability type $l$, where $1\leq l \leq m=40$.

The metrics for evaluating multiclass vulnerability detectors are
$M\_FPR$ (the multiclass counterpart of the false positive rate based on the total number of vulnerability classes),
$M\_FNR$ (the multiclass counterpart of the false negative rate based on the total number of vulnerability classes),
$M\_F1$ (the multiclass counterpart of the F1 measure based on the total number of vulnerability classes),
$W\_FPR$ (the multiclass counterpart of the false positive rate based on the weight of each vulnerability class),
$W\_FNR$ (the multiclass counterpart of the false negative rate based on the weight of each vulnerability class),
$W\_F1$ (the multiclass counterpart of F1 measure based on the weight of each vulnerability class). The $M\_*$ metrics are arithmetic average of the metrics of the overall vulnerability classes. The $W\_*$ metrics are the average in which metrics of each vulnerability class are multiplied by the weight before summing. The weight here refers to the proportion of the number of each vulnerability class in the total number of vulnerabilities.
Specifically, we have:
\begin{eqnarray*}
M\_FPR &=& \frac{1}{\left | L\right |}\begin{matrix} \sum_{l\in L} \frac{FP_l}{FP_l+TN_l}    \end{matrix},\\
M\_FNR &=& \frac{1}{\left | L\right |}\begin{matrix} \sum_{l\in L} \frac{FN_l}{TP_l+FN_l}    \end{matrix},\\
M\_F1 &=& \frac{1}{\left | L\right |}\begin{matrix} \sum_{l\in L}  \frac{2\cdot P_l\cdot R_l}{P_l+R_l}  \end{matrix},\\
W\_FPR &=&
\begin{matrix}
\frac{1}{\sum_{l\in L} \left | X_l\right |}\sum_{l\in L} \left | X_l\right |\cdot \frac{FP_l}{FP_l+TN_l}
\end{matrix},\\
W\_FNR &=&
\begin{matrix}
\frac{1}{\sum_{l\in L} \left | X_l\right |}\sum_{l\in L}\left | X_l\right |\cdot \frac{FN_l}{TP_l+FN_l}
\end{matrix},\\
W\_F1 &=&
\begin{matrix}
\frac{1}{\sum_{l\in L} \left | X_l\right |} \sum_{l\in L}\left | X_l\right |\cdot \frac{2\cdot P_l\cdot R_l}{P_l+R_l}
\end{matrix}.
\end{eqnarray*}
Note that when $|L|=1$, $M\_FPR$ naturally degenerates to the false positive rate ($FPR$) of binary vulnerability detector,
$M\_FNR$ naturally degenerates to the false negative rate ($FNR$) of binary vulnerability detector, and
$M\_F1$ naturally degenerates to the $F1$ measure of binary vulnerability detector.
However, $W\_FPR$, $W\_FNR$, and $W\_F1$ do not have their counterparts when $|L|=1$.

%% file: experiment.tex
Our experiments are centered at answering the following Research Questions (RQs):
\begin{itemize}	
\item RQ1: how effective is $\mu$VulDeePecker for multiclass vulnerability detection?
\item RQ2: does accommodating control-dependence indeed improve the effectiveness of $\mu$VulDeePecker in multiclass vulnerability detection?
\item RQ3: can $\mu$VulDeePecker be used with other deep-learning-based systems (e.g., VulDeePecker) to obtain higher effectiveness?
\end{itemize}

Whenever appropriate, we will consider the following extension of VulDeePecker (which was designed for detecting whether a program is vulnerable or not), dubbed VulDeePecker+, for multiclass vulnerability detection:
Unlike VulDeePecker, which uses the sigmoid activation function and the binary crossentropy loss function, we propose using the softmax activation function and the categorical crossentropy loss function, leading to a variant which we call VulDeePecker+.
We should compare $\mu$VulDeePecker against with other multiclass detectors. Since open-source tools (e.g., Flawfinder) have poor detection capabilities \cite{li2018vuldeepecker} and we have no access to commercial tools (e.g., Fortify and Coverity, no budget to buy), we only compare $\mu$VulDeePecker against VulDeePecker+.
Our experiments are performed on a computer with an Intel Xeon E5-1620 CPU operating at 3.50GHz and an NVIDIA GeForce GTX 1080 GPU. The operating system is Linux 3.10.0-514.6.2.el7.x86\_64.

\subsection{Experiments for Answering RQ1}
\label{sec:Evaluation on VulDeePecker+}

In order to evaluate the effectiveness of $\mu$VulDeePecker, we apply it to (i) the testing set mentioned above and (ii) real-world product software.

\subsubsection{Experiments on Testing Set}

Table \ref{t:fine} summarizes the experimental results.
We observe that M\_FPR, M\_FNR, and M\_F1 of $\mu$VulDeePecker are 0.02\%, 5.73\%, and 94.22\%, respectively.
When compared with VulDeePecker+, $\mu$VulDeePecker is 0.01\% lower in terms of M\_FPR, 10.75\% lower in terms of M\_FNR, and 8.72\% higher in terms of M\_F1.
This might be explained by the fact that the local features learned from code attentions accommodate much information about vulnerability types, leading to much smaller multiclass false negative rate.
Moreover, measurements of W\_FPR, W\_FNR, and W\_F1 also indicate that $\mu$VulDeePecker is more effective than VulDeePecker+. As to the test time, although the time of $\mu$VulDeePecker is slightly higher than VulDeePecker+, considering the detection effectiveness of $\mu$VulDeePecker, the time consumption is tolerable.

\begin{table}[!htbp]
	\centering
	\small	
	\caption{Experimental results on the testing set with respect to $\mu$VulDeePecker and VulDeePecker+ (for answering RQ1).}
	\begin{tabular}{m{1.cm}|m{0.65cm}|m{0.68cm}|m{0.6cm}|m{0.65cm}|m{0.68cm}|m{0.6cm}|m{0.7cm}}
		\toprule
		Models&M\_FP R(\%)&M\_FN R(\%)&M\_F1 (\%)&W\_FP R(\%)&W\_FN R(\%)&W\_F1 (\%)& Time (s)\\
		\hline
		$\mu$VulDe-ePecker & \textbf{0.02}& \textbf{5.73}& \textbf{94.22} & \textbf{0.13}& \textbf{7.49}& \textbf{94.69} &1413.12\\
		\hline
		VulDee-Pecker+ & 0.03& 16.48& 85.5& 0.3& 7.37& 93.61&1063.14\\
		\bottomrule
	\end{tabular}
	\label{t:fine}	
\end{table}

\begin{figure*}[!htbp]
	\centering	
	\includegraphics[width=\textwidth]{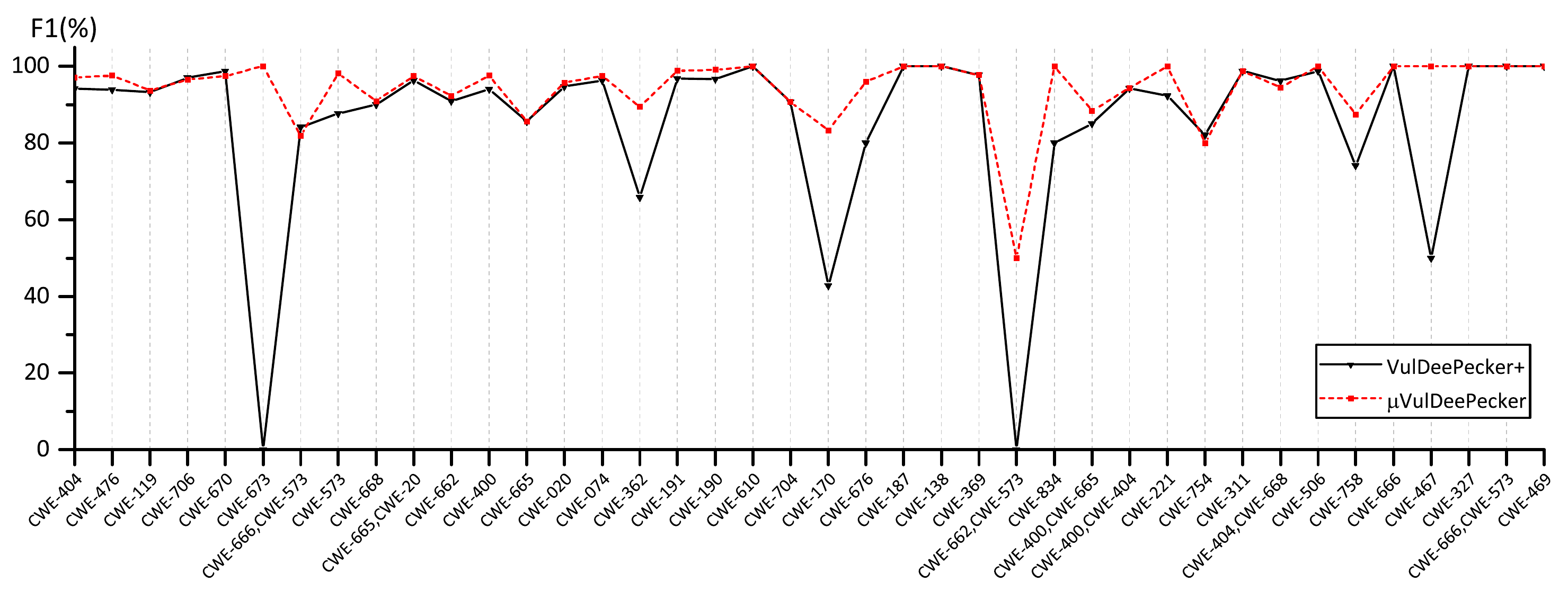}
	\caption{Experimental comparison between the effectiveness of $\mu$VulDeePecker and VulDeePecker+ with respect to each of the 40 vulnerability types.}
	\label{fig:mode_40}
\end{figure*}

Fig \ref{fig:mode_40} plots the detection result of $\mu$VulDeePecker and VulDeePecker+, in terms of the F1, with respect to each of the 40 vulnerability types.
We observe that $\mu$VulDeePecker is overall more effective than VulDeePecker+, especially for
vulnerability types CWE-673, CWE-362, CWE-170, sub-type of CWE-662 and CWE-573, and CWE-467.
In order to explain this discrepancy, we observe that these five types have small numbers of vulnerabilities: 16 vulnerabilities for CWE-673, 211 for CWE-362, 45 for CWE-170, 33 for the sub-type of CWE-662 and CWE-573, and 55 for CWE-467; these vulnerabilities might have been ignored by VulDeePecker+ as noise.
For example, CWE-467 is a type of vulnerabilities related to function call $sizeof()$,
which cannot be recognized by VulDeePecker+ because it often occurs in conjunction with buffer overflows (i.e., another type of vulnerabilities).
In contrast, $\mu$VulDeePecker can cope with $sizeof()$ by leveraging its arguments, which are captured by code attentions.

\subsubsection{Experiments on Real-World Software}

We apply $\mu$VulDeePecker to 10 versions of 3 real-world software products, namely Libav, Seamonkey, and Xen. Since we do not know whether these products contain vulnerabilities or not (i.e., the ground truth is not available), we manually examine and confirm the detected vulnerabilities. Experimental results show that $\mu$VulDeePecker detects 16 vulnerabilities, 
among which 14 vulnerabilities correspond to patterns of known vulnerabilities.
Among these 14 vulnerabilities, 6 vulnerabilities (respectively corresponding to the patterns of CVE-2013-0866, CVE-2013-4264, CVE-2013-7012, CVE-2013-7022, CVE-2013-7023, and CVE-2014-9319) are detected in Libav, 3 vulnerabilities (respectively corresponding to the patterns of CVE-2015-4511, CVE-2015-4513, and CVE-2015-4517) are detected in Seamonkey, and 
5 vulnerabilities (respectively corresponding to the patterns of CVE-2013-4149, CVE-2013-4150, CVE-2014-5263, CVE-2016-4952, and CVE-2016-9923) are detected in Xen. These vulnerabilities include out-of-bounds read, use-after-free, and improper restriction of operations within the bounds of a memory buffer.
The other 2 vulnerabilities are new because they are not known to exist in Xen 4.4.0 to the public until now (despite that they respectively correspond to the patterns of CVE-2015-7512 and CVE-2016-5126), but have been silently fixed by the software vendor when releasing newer versions.
These 2 vulnerabilities are caused by buffer overflow and are highlighted in Table \ref{t:real} (for ethical reasons, we anonymized the file name of the vulnerable program). We map CVE\# to them in Table \ref{t:real} because they were detected via the patterns corresponding to those CVE\#'s. Summarizing preceding discussions, we draw:

\begin{table*}[h]
	\centering
	\small	
	\caption{The 2 vulnerabilities, whose existence is not known until now, found by $\mu$VulDeePecker in real-word software products.}
	\begin{tabular}{m{3cm}|m{2.2cm}|m{1.5cm}|m{4cm}|m{2cm}|m{2.5cm}}
		\toprule
		Real-world software&CVE-ID &CWE-ID &The path of vulnerable files&Library/API
		function call&The first patched version\\
		\hline
		Xen 4.4.0 & CVE-2015-7512& CWE-119& .../hw/net/{\tt anonymized}.c &memcpy& Xen 4.6.3\\
		\hline
		Xen 4.4.0 & CVE-2016-5126& CWE-119& .../block/{\tt anonymized}.c & memcpy& Xen 4.8.0\\
		\bottomrule
	\end{tabular}
	\label{t:real}	
\end{table*}

\begin{insight}
$\mu$VulDeePecker is effective for multiclass  vulnerability detection, especially its use of local features learned from code attentions makes it effective even when samples are small.
\end{insight}

\subsection{Experiments for Answering RQ2}

In order to quantify the value of control-dependence in multiclass vulnerability detection, we prepare a dataset that only contains data-dependence, which is indeed obtained by considering data-dependence relation only in Step I.1 of $\mu$VulDeePecker. The same 80\% of the dataset (i.e., the code gadgets corresponding to the same programs in the training set for learning $\mu$VulDeePecker) is used for training, and the rest 20\% is used for testing.

\begin{table}[!htbp]
	\centering
	\small	
	\caption{Experimental results of $\mu$VulDeePecker with respect to different datasets (for answering RQ2).}
	\begin{tabular}{m{1.1cm}|m{0.85cm}|m{0.92cm}|m{0.6cm}|m{0.85cm}|m{0.92cm}|m{0.5cm}}
		\toprule
		Models&M\_FPR (\%)&M\_FNR (\%)&M\_F1 (\%)&W\_FPR (\%)&W\_FNR (\%)&W\_F1 (\%)\\
\hline
		\multicolumn{7}{c}{Dataset accommodating data and control dependences}\\
		\hline
		$\mu$VulDe-ePecker & \textbf{0.02}& \textbf{5.73}& \textbf{94.22} & \textbf{0.13}& \textbf{7.49}& \textbf{94.69}\\
		\hline
		\multicolumn{7}{c}{Dataset accommodating data dependence only}\\
		\hline
		$\mu$VulDe-ePecker & 0.1& 18& 81.59& 0.94& 14.39& 83.56\\
		\bottomrule
	\end{tabular}
	\label{t:mulRQ4}	
\end{table}

Table \ref{t:mulRQ4} summarizes the experimental results.
We observe that when additionally accommodating control-dependence, $\mu$VulDeePecker can improve M\_F1 by 12.63\%, W\_F1 by 11.13\% and decrease M\_FNR by 12.27\%, W\_FNR by 6.9\%.
This justifies the value of accommodating control-dependence relation in multiclass vulnerability detection, which is intuitive. In summary, we draw:

\begin{insight}
Accommodating control-dependence enhances the capability of $\mu$VulDeePecker in multiclass vulnerability detection.
\end{insight}

\subsection{Experiments for Answering RQ3}
\label{sec:Evaluation on datasets}

In order to answer RQ3, we use $\mu$VulDeePecker and VulDeePecker+ together with respect to the dataset accommodating both data-dependence and control-dependence, as follows: For a code gadget sample (including its code attention), suppose $\mu$VulDeePecker detects it as type-$i$ with probability $p_i$ and VulDeePecker+ detects it as type-$j$ with probability $q_j$, where $0\leq i, j \leq m$ (with $m=40$ corresponding to our dataset).
Then, the vulnerability type of the sample is set to be $i$ or $j$ corresponding to the maximum probability, namely
$$\max(p_i,q_j)=\max(p_0,\ldots,p_m,q_0,\ldots,q_m).$$

Table \ref{t:bin_rq3} summarizes the results, which show that using $\mu$VulDeePecker and VulDeePecker+ together indeed leads to better detection accuracy (2.65\% and 1.59\% higher in M\_F1 and W\_F1).
This suggests that $\mu$VulDeePecker and VulDeePecker+ accommodate different kinds of information useful for multiclass vulnerability detection. In summary, we draw:

\begin{table}[!htbp]
	\centering
	\small
	\caption{Experimental results with respect to use one detector or multiple detectors (for answering RQ3).}
	\begin{tabular}{m{1.1cm}|m{0.85cm}|m{0.92cm}|m{0.6cm}|m{0.85cm}|m{0.92cm}|m{0.5cm}}
		\toprule
		Models&M\_FPR (\%)&M\_FNR (\%)&M\_F1 (\%)&W\_FPR (\%)&W\_FNR (\%)&W\_F1 (\%)\\
		\hline
		$\mu$VulDe-ePecker & 0.02& 5.73& 94.22 & 0.13& 7.49& 94.69\\
		\hline
		using both & \textbf{0.01}& \textbf{4.46}& \textbf{96.87} & \textbf{0.08}& \textbf{5.53}& \textbf{96.28}\\
		\bottomrule
	\end{tabular}
	\label{t:bin_rq3}
\end{table}

\begin{insight}
$\mu$VulDeePecker can be used together with other deep-learning-based systems such as VulDeePecker+ to accommodate more, useful information for multiclass vulnerability detection.
\end{insight}